# Selective etching of PDMS: etching as positive resist

*S.Z. Szilasi*[1,*], *C. Cserháti*[2]

[1] *Institute for Nuclear Research, Hungarian Academy of Sciences,*

*H-4001 Debrecen, P.O. Box 51, Hungary*

[2] *University of Debrecen, Dept. of Solid State Physics,*

*H-4010 Debrecen, P.O. Box 2, Hungary*

\* Corresponding author. S.Z. Szilasi

Address: Institute of Nuclear Research of the Hungarian Academy of Sciences

H-4026 Debrecen Bem tér 18/c, Mail: H-4001 Debrecen, POB. 51 Hungary

E-mail address: szilasi.szabi@gmail.com

Tel: +36 52 509 200; Fax: +36 52 416 181

**Abstract**

Although, poly(dimethylsiloxane) (PDMS) is a widely used material in numerous applications, such as micro- or nanofabrication, the method of its selective etching has not been known up to now. In this work authors present two methods of etching the pure, additive-free and cured PDMS as a positive resist material.

To achieve the chemical modification of the polymer necessary for selective etching, energetic ions were used. We created 7 μm and 45 μm thick PDMS layers and patterned them by a focused proton microbeam with various, relatively large fluences. In this paper authors demonstrate that 30 wt% Potassium Hydroxide (KOH) or 30 wt% sodium hydroxide (NaOH) at 70 ºC temperature etch proton irradiated PDMS selectively, and remove the chemically sufficiently modified areas. In case of KOH development, the maximum etching rate was approximately 3.5 μm/minute and it occurs at about $7.5\times10^{15}$ ion × cm$^{-2}$. In case of NaOH etching the maximum etching rate is slightly lower, 1.75 μm/minute and can be found at the slightly higher fluence of $8.75\times10^{15}$ ion × cm$^{-2}$.

These results are of high importance since up to this time it has not been known how to develop the additive-free, cross-linked poly(dimethylsiloxane) in lithography as a positive tone resist material.

1. Introduction

The continuous development of lithographic techniques demand the improvement of both the development methods and resist materials. By emerging new methods or materials, new possibilities become attainable making the existing methods simpler, more reliable, better quality or faster.

Poly(dimethylsiloxane) (PDMS) is the most widely used silicon-based, organic, cross-linkable polymer. The cross-linked PDMS is a rubbery solid, it does not permanently deform under stress or strain. Since it is chemically inert and by a chemical or physical treatment it can be turned into biocompatible, PDMS is widely used in medical fields also. Its high oxygen permeability, good mechanical properties, chemical stability and easy processing make it an ideal raw material for ophthalmological products (e.g. contact lenses [1]) or implants that are in direct, and sometimes prolonged contact with human tissues [2]. The high optical clarity, low attenuation [3] and the excellent stability against weathering makes this polymer applicable for creating optical waveguides or microlenses [4,5]. Poly(dimethylsiloxane) is undoubtedly the most commonly used microfluidic material in research laboratories. It is also hydrophobic, chemically resistive, cost effective and easy to use. With plasma treatment, it can easily be bonded to another PDMS layer, to a glass or Si substrate [6]. All these features together are highly desirable at fabrication and prototyping of microfluidic chips and lab-on-a-chip devices. These devices, and thus Poly(dimethylsiloxane), have demonstrated significant potential in countless applications, such as chemical separation [7], separation and processing of biological cells [8], fuel cells [9], chemical microreactors [10] or even spacecraft thrusters [11].

Despite its versatility and the numerous advantageous properties mentioned above, PDMS is mainly used as a mold, a casting or replicating material [12]. However, recent researches pointed out that PDMS can be applied as a resist material also. In 2002, Constantoudis et.al. applied electron beam lithography to pattern a thin layer of liquid, uncured PDMS and then used the structures as a hard mask [13]. In 2009, Szilasi et.al. observed that significant compaction occurs at the irradiated areas [14] and applied it to create parallel lines with curved surfaces [15] and microlenses [4] in one step, without the need of any further development. In 2011, Tsuchiya et.al. reported that the uncured, liquid phase PDMS polymer crosslinks, thus acts as a negative resist if it is exposed to proton irradiation and created microstructures in it [16]. Bowen et.al. created structures by electron beam lithography and studied the change of Young's modulus as a function of the delivered dose in 2012 [17]. In 2016, Gorissen et.al. patterned PDMS through SU8 mask by reactive ion etching (RIE) [18]. Others use various additives to make the pre-polymer photosensitive [19] and apply photolithography.

The key to the application of PDMS as a resist material is the change of its chemical properties where it is exposed to various types of radiation. Due to irradiation, chain scissioning happens in the polymer, the main Si-O-Si chain brakes, functional groups split and the volatile products (e.g. $H_2$, $CH_4$ and $C_2H_6$ gases) leave the irradiated volume [20,21]. As a result of these processes, PDMS shrinks at the place of irradiation and the initially elastic polymer becomes a rigid, brittle and glass-like material [15]. Since the cured PDMS is chemically very resistant, it has been unknown how to selectively etch the modified areas to create micro/nanostructures directly in the polymer.

Some studies show that cured and unirradiated PDMS resists potassium hydroxide (KOH) solution relatively well. Brugger et.al. used PDMS seal rings during etching of silicon wafers with KOH solution at 60 ºC [22]. In these experiments, the PDMS rings were reused many times after more than 30 hours of etching. Mata et.al. found that shallow PDMS microstructures damage after immersing them into KOH solution for 27 hours [23].

The above result show that cured poly(dimethylsiloxane) withstands KOH solution for an extended period of time, but it is also known that amorphous silicon, crystalline silicon and $SiO_2$ can be selectively etched by the same solution [24, 25]. Since the irradiation degrades the polymer significantly which becomes glass-like, and $SiO_x$ forms [21], we expected that the reaction rate of KOH with the degraded PDMS would increase. In many applications potassium hydroxide is almost interchangeable with sodium hydroxide (NaOH). The two substances have very similar chemical properties, so we tested and compared the performance of both in etching irradiated PDMS. The result of this study is the subject of this paper.

Since the method of selective etching of PDMS as a positive tone resist has not been known up to know, these results may be of high importance in various micro- and nanolithography techniques. Our findings make possible the direct / maskless creation of micro- and nanostructures, microfluidic systems or even lithography masks in fewer steps, without the need of molds, directly in cured PDMS.

## 2. Material and methods

To create the samples, Sylgard 184 kit (Dow-Corning) was used. The base polymer and the curing agent were mixed with the volume ratio of 10:1, respectively, spin coated on glass substrates in 7 μm and 45 μm thicknesses and baked at 125 °C for 30 min.

The samples were irradiated by a 2 MeV focused proton beam at the nuclear microprobe facility at HAS-ATOMKI, Debrecen, Hungary [26]. The size of the beam spot was 2.5 μm × 2.5 μm, the scanning resolution of the irradiated patterns (i.e. the distance between neighbouring pixels) was ~1 μm. Since the beam spot was larger than the scanning resolution, the beam spot overlaps with itself multiple times when scanning neighbouring pixels. SRIM [27] calculations showed that the penetration depth for 2 MeV protons is ~85 μm in PDMS. Since the polymer layer is much thinner than the maximal penetration depth, the particles easily penetrate through the resist layer without considerable lateral scattering making the creation of vertical sidewalls possible. To test if the etching method works and to find the ideal irradiation and etching parameters, 7 μm thick samples were irradiated with 15 parallel lines. Each line received a different fluence in increasing order between $1.33 \times 10^{15}$ ion × cm$^{-2}$ (2130 nC×mm$^{-2}$) and $2 \times 10^{16}$ ion × cm$^{-2}$ (32 050 nC×mm$^{-2}$) in approximately $1.25 \times 10^{15}$ ion × cm$^{-2}$ (2000 nC×mm$^{-2}$) increments. These samples were called fluence test samples. Other test structures, such as circles with different diameters, squares in different sizes and lines with various widths were also irradiated in 7 μm and 45 μm thick cured PDMS. These structures received $1.19 \times 10^{16}$ ion × cm$^{-2}$ (19 000 nC×mm$^{-2}$) fluence. The beam current was 1.4 nA in every case.

To remove the irradiated areas of the cured PDMS, 30 wt% Potassium Hydroxide (KOH) and 30 wt% sodium hydroxide (NaOH) were used. The samples were placed in one of these solutions for various times to find the best etching parameters. Since the temperature of the solutions was 70 °C in every case, the beaker needed to be covered to avoid evaporation and thus the change of the concentration of the etchant. The solutions were continuously stirred magnetically during etching.

The decontamination of the structures following etching was done by immersing them into 5:1:1 $H_2O:H_2O_2:HCl$ solution for 2 minutes and then they were rinsed in distilled water.

The etched structures were investigated by optical microscopy and scanning electron microscopy (SEM).

## 3. Results and Discussion

A fluence test sample, that was irradiated with fifteen parallel lines and each line received different fluences, was attempted to etch at room temperature in 30 wt% KOH and 30 wt% NaOH. The etchant was stirred continuously and the progress of etching was recorded in every 5 minutes. After 50 minutes, no etching was observed at any fluences in either etchants.

This experiment was repeated at elevated temperatures. The observations show that in this case both KOH and NaOH etches effectively the irradiated areas of the cured PDMS while the unirradiated parts do not etch away. This indicates that the etching rate of the irradiated PDMS is highly dependent of the temperature of the KOH and NaOH solutions. The etching in both cases was found to be isotropic and homogenous.

### 3.1 Etching with Potassium Hydroxide (KOH)

A fluence test sample was taken out from the KOH solution after certain period of times. After each time, the advancement of the etching process have been investigated. The cured PDMS is quite resistant to the KOH solution, while the irradiated areas etch quickly. After 2 minutes of etching, the lines with $5.31 \times 10^{15}$ ion $\times$ cm$^{-2}$ (8 500 nC$\times$mm$^{-2}$), $6.53 \times 10^{15}$

ion × cm$^{-2}$ (10 450 nC×mm$^{-2}$), 8.02×10$^{15}$ ion × cm$^{-2}$ (12 830 nC×mm$^{-2}$), 9.09×10$^{15}$ ion × cm$^{-2}$ (14 550 nC×mm$^{-2}$) fluences were etched completely to the substrate (Figure 1). Higher magnification images show, that the lines with 5.31×10$^{15}$ ion × cm$^{-2}$ (8 500 nC×mm$^{-2}$) and 9.09×10$^{15}$ ion × cm$^{-2}$ (14 550 nC×mm$^{-2}$) fluences still had some glassy remnants inside or at the edge of the irradiated lines. The cleanest structure after 2 minutes of etching was the line with 8.02×10$^{15}$ ion × cm$^{-2}$ (12 830 nC×mm$^{-2}$) fluence.

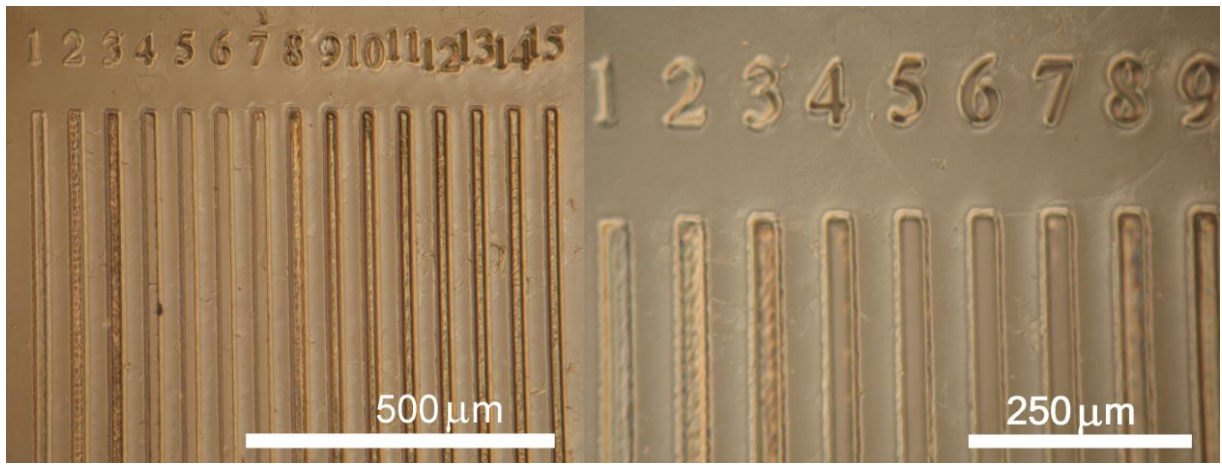

**Figure 1. After 2 minutes of etching, the lines with 5.31×10$^{15}$ ion × cm$^{-2}$ (8 500 nC×mm$^{-2}$, #4), 6.53×10$^{15}$ ion × cm$^{-2}$ (10 450 nC×mm$^{-2}$, #5), 8.02×10$^{15}$ ion × cm$^{-2}$ (12 830 nC×mm$^{-2}$, #6), 9.09×10$^{15}$ ion × cm$^{-2}$ (14 550 nC×mm$^{-2}$, #7) fluences were etched completely to the substrate. The cleanest structure was the one with 8.02×10$^{15}$ ion × cm$^{-2}$ (12 830 nC×mm$^{-2}$, #6) fluence.**

As it was shown earlier, the degradation of poly(dimethylsiloxane) increases with increasing fluences [21]. Considering this, the above observation is surprising because it means that the etching rate of the irradiated PDMS does not change according to the degree of degradation of the polymer matrix. There is apparently an optimal change in the structure of the polymer that is needed to achieve the highest reaction rate with KOH and thus the highest etching speed. This optimal change in the structure of the polymer can be assigned to a certain fluence. Above this fluence, the more degraded and more glass-like polymer structure results lower etching rates. To clarify what the reason of this phenomenon is, further chemical and structural investigations of the irradiated PDMS will be needed. The results show that the fastest etching occurs at around the fluence of 7.5×10$^{15}$ ion × cm$^{-2}$ (12 000 nC×mm$^{-2}$), below and above this fluence the etching rate is lower.

After the above observations, the etching was continued. The sample was placed back into KOH and after every minute the advancement of the etching was recorded. The results show, that the structures with different fluences needed different times to etch completely. The calculated etching rates in the function of the delivered charge densities are shown in Table 1 and Figure 2.

| Line number | Fluence (x10$^{15}$ ion/cm²) | Charge densities (nC/mm²) | Time needed to etch (minutes) | Etching rate (μm/minute) |
|---|---|---|---|---|
| 1 | 1.33 | 2130 | | |
| 2 | 2.74 | 4380 | 10 | 0.70 |
| 3 | 3.99 | 6390 | 4 | 1.75 |
| 4 | 5.31 | 8500 | 3 | 2.33 |
| 5 | 6.53 | 10450 | 2 | 3.50 |
| 6 | 8.02 | 12830 | 2 | 3.50 |
| 7 | 9.09 | 14550 | 3 | 2.33 |
| 8 | 10.70 | 17090 | 4 | 1.75 |
| 9 | 11.90 | 19100 | 5 | 1.40 |
| 10 | 13.90 | 22220 | 6 | 1.17 |
| 11 | 14.80 | 23670 | 7 | 1.00 |
| 12 | 16.50 | 26360 | 8 | 0.88 |
| 13 | 17.60 | 28170 | 9 | 0.78 |
| 14 | 19.00 | 30450 | 11 | 0.64 |
| 15 | 20.00 | 32050 | 12 | 0.58 |

**Table 1. Etching irradiated PDMS with KOH solution: times needed to etch through a 7 μm thick layer and the calculated etching rates at the different delivered fluences**

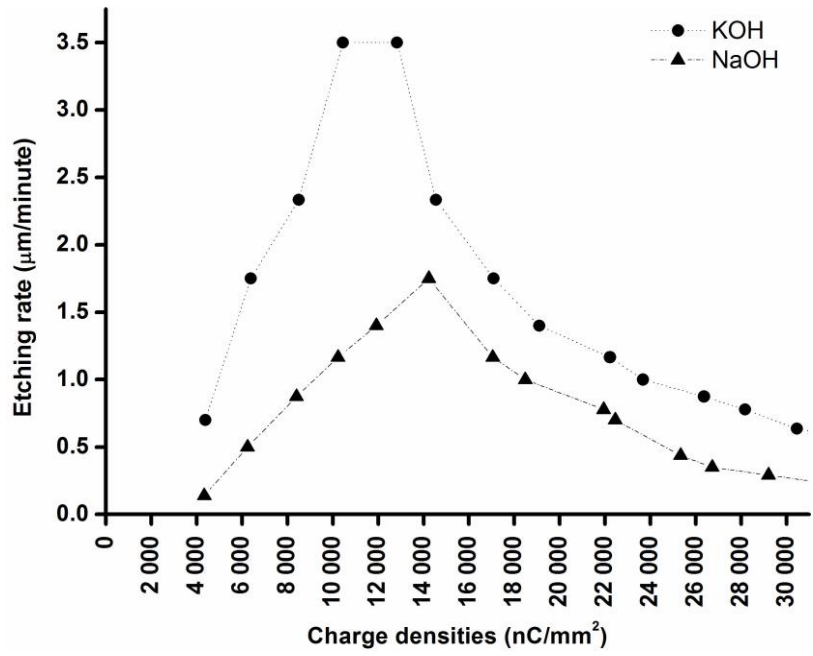

**Figure 2.** Etching rate of irradiated PDMS in the function of charge densities. The maximum etching rate in case of KOH development is at about 12 000 nC×mm$^{-2}$ (7.5×10$^{15}$ ion × cm$^{-2}$), while in case of NaOH etching it is at 14 000 nC×mm$^{-2}$ (8.75×10$^{15}$ ion × cm$^{-2}$).

In case of the structures that were irradiated with more than $8.02\times10^{15}$ ion × cm$^{-2}$ (12 830 nC×mm$^{-2}$) fluence, the etching occurred faster along the edges of the structures than the inside parts (Figure 3.). The reason for this is that the beam spot is larger than the lateral resolution of the irradiation, so during scanning neighbouring pixels the beam spot overlaps multiple times. The number of overlappings is less along the edges of the structure, so these areas receive lower fluence. As shown in Figure 2, above approximately $7.5\times10^{15}$ ion × cm$^{-2}$ (12 000 nC×mm$^{-2}$) fluence the etch rate decreases with increasing fluences. Since the fluence is lower along the edges of the structures, these areas etch faster than the inside parts.

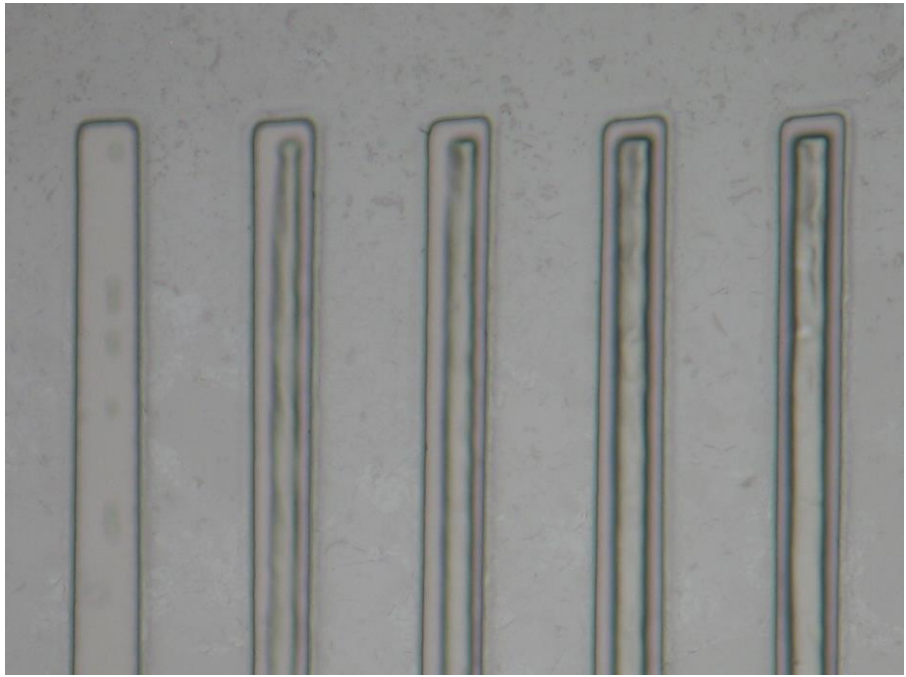

**Figure 3.**
**As the fluence increases above $7.5\times10^{15}$ ion × cm$^{-2}$ (12 000 nC×mm$^{-2}$), the etching speed decreases and the structures etch slower. However, the edge of the high fluence structures etch faster than their inside parts because the delivered dose is lower along the edges due to the fewer overlappings of the beam spot.**

Other samples were also created with different test structures, such as various diameter circles, dots matrices, squares and parallel lines (Figure 4).

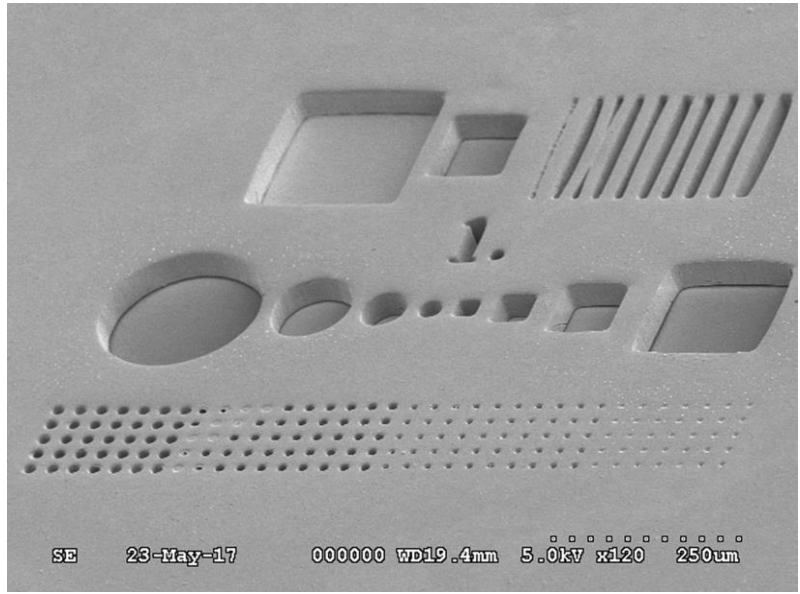

**Figure 4. Test structures etched by KOH in 45 μm thick cured PDMS**

These structures received $1.19 \times 10^{16}$ ion × cm$^{-2}$ (19 000 nC×mm$^{-2}$) fluence where the etching rate is ~1.4 μm/minute. The sample was 45 μm thick, the etching lasted for 40 minutes to achieve the best result. The structures developed in good quality, only some small dimension ones did not etch completely (such as a line with the width of 1 pixel and the single pixel dots). The problem in these cases was probably that the etching liquid could not circulate properly inside the small cavities and thus they were etched much slower. To enhance the etching of these small structures, ultrasonic agitation was applied. Unfortunately, the agitation that lasted longer than few seconds caused tear at the edges of some structures and the PDMS layer started to peel off.

During the development experiments, it was found that the 30 wt% KOH + 20 wt% IPA + 50 wt% DI water solution at 70 °C temperature etched away both the irradiated and non-irradiated PDMS in 20 minutes. This solution can be used to clean any PDMS residues off of glass wafers.

It was also found that 98% sulfuric acid etches away the unirradiated PDMS while it keeps the irradiated areas, making thus the etching of PDMS as a negative tone resist possible. This finding is studied in details in a separate paper.

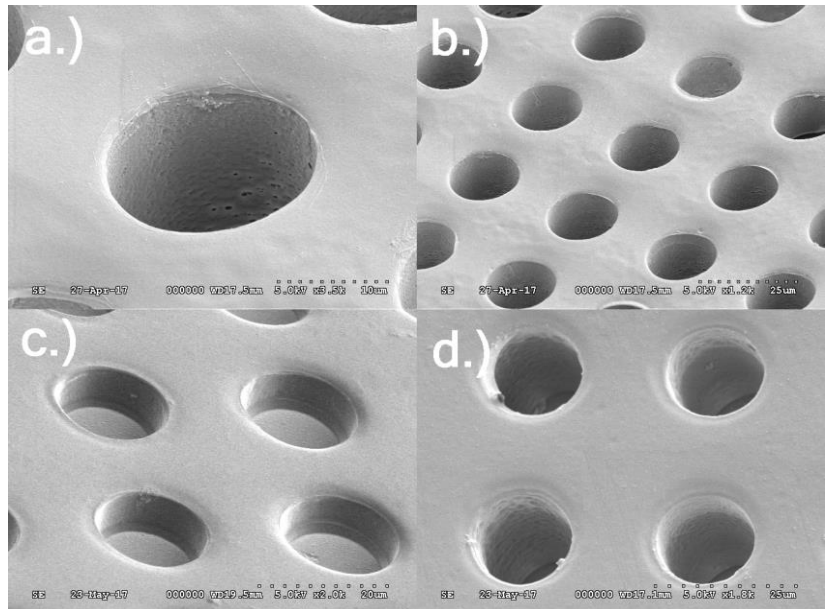

**Figure 5. 18 μm in diameter holes are etched with KOH to the substrate in 7 μm (c.) and 45 μm (a.,b.,d.) thick PDMS**

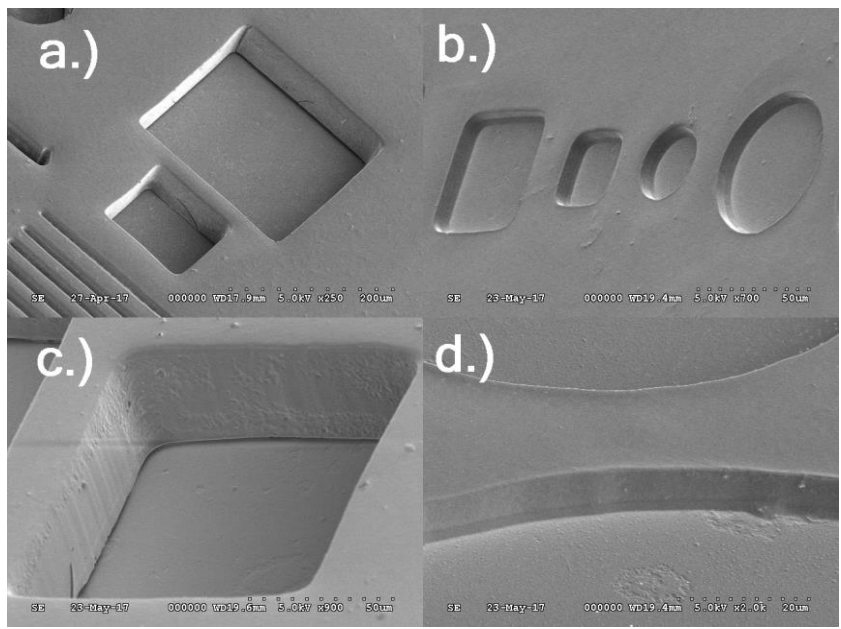

**Figure 6. Structures etched with KOH to the substrate in 45 μm (a., c.) and 7 μm (b., d.) thick PDMS**

### 3.2 Etching with sodium hydroxide

Since sodium hydroxide and potassium hydroxide has similar chemical properties, the etching behaviour of NaOH was expected to be similar to that of KOH. The development procedure of the fluence test sample was the same as in case of KOH etching, the results are summarized in Table 2. and Figure 2.

| Line number | Fluence ($\times 10^{15}$ ion/cm$^2$) | Charge densities (nC/mm$^2$) | Time needed to etch (minutes) | Etching rate (µm/minute) |
|---|---|---|---|---|
| 1 | 1.33 | 2130 | | |
| 2 | 2.70 | 4330 | 50 | 0.14 |
| 3 | 3.90 | 6240 | 14 | 0.50 |
| 4 | 5.25 | 8400 | 8 | 0.88 |
| 5 | 6.39 | 10230 | 6 | 1.17 |
| 6 | 7.45 | 11920 | 5 | 1.40 |
| 7 | 8.89 | 14240 | 4 | 1.75 |
| 8 | 10.65 | 17050 | 6 | 1.17 |
| 9 | 11.54 | 18480 | 7 | 1.00 |
| 10 | 13.70 | 21940 | 9 | 0.78 |
| 11 | 14.02 | 22450 | 10 | 0.70 |
| 12 | 15.82 | 25330 | 16 | 0.44 |
| 13 | 16.69 | 26720 | 20 | 0.35 |
| 14 | 18.24 | 29200 | 24 | 0.29 |
| 15 | 19.78 | 31670 | 30 | 0.23 |

**Table 2. Etching irradiated PDMS with NaOH: times needed to etch through a 7 µm thick layer and the calculated etching rates in the function of the delivered fluence**

As it can be seen in Figure 2., the etch rate curve for NaOH has similar character to that of KOH, but the etching rates are about 1.5 - 2 times lower at every fluences. The highest etch rate was found at about $8.89 \times 10^{15}$ ion $\times$ cm$^{-2}$ (14 240 nC$\times$mm$^{-2}$) fluence, which was a little higher than in case of KOH. Since the etching was slower above this fluence, similarly to the case of KOH etching, higher fluence samples etched faster along its edges. SEM images showed that the smoothness and quality of the sidewalls improved with increasing fluences.

The test structures, that received $1.19 \times 10^{16}$ ion $\times$ cm$^{-2}$ (19 000 nC$\times$mm$^{-2}$) fluence, were developed in good quality (Figure 7.). Although, the single pixel wide figures were not etched completely.

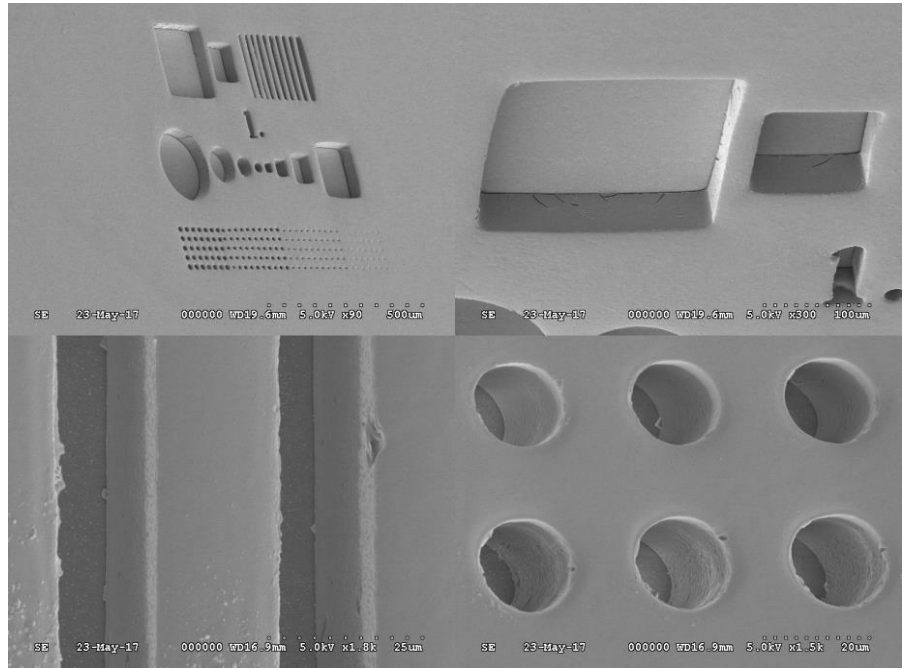

**Figure 7 - Structures created in 45 μm thick PDMS by proton irradiation and etched by NaOH.**

## 4. Conclusions

In this work, we are presenting a method how irradiated PDMS can be selectively etched as a positive tone resist material. Because PDMS is easy to handle, it is widely used as a negative resist material in electron beam lithography as well as in proton beam lithography to create different patterns and to prototype different hard masks for special applications. Since the application of PDMS as a positive tone resist has numerous potential advantages in micro- and nanopatterning, lots of efforts were done to develop a method that makes this possible.

In this study, PDMS was irradiated with 2 MeV protons with various fluences and the irradiated areas were etched with KOH and NaOH solutions. The concentration of KOH and NaOH was 30 wt% in every cases, their temperature during etching was always 70 ºC. In case of KOH etching, the maximal etching rate occurred at about $7.5\times10^{15}$ ion × cm$^{-2}$ (12 000 nC×mm$^{-2}$) proton fluence and it was approximately 3.5 µm/minute. In case of NaOH, the maximal etching rate is slightly lower, 1.75 µm/minute and it can be found at about $8.89\times10^{15}$ ion × cm$^{-2}$ (14 240 nC×mm$^{-2}$) fluence. The etching at room temperature happens much slower, after 50 minutes no etching was observed at any fluences in either of the etchants. This shows that the etching rate strongly depends on the temperature of the etchant.

It was also observed that the smoothness and quality of the sidewalls improves with increasing fluences.

In case of those structures that received more fluence than the value where the maximal etching rate occurs, the edges of the structures etch faster than their inside parts. This is caused by the overlappings of the beam spot. Since the dimension of the beam spot is larger than the lateral resolution of the irradiation, so during scanning neighbouring pixels the beam spot overlaps multiple times. The number of overlappings is less along the edges of the structures, so these areas receive lower fluence. Since at high doses the etching rate increases with decreasing fluences, the edges will etch faster than the inside parts of the structures.

Various test structures were created with $1.19\times10^{16}$ ion × cm$^{-2}$ (19 000 nC×mm$^{-2}$) fluence in 7 µm and 45 µm thick PDMS polymer and etched with both etchants. The structures etched with good quality down to the substrate, only the smallest ones did not develop completely. The reason for this probably was that the etchant could not circulate adequately inside small cavities and thus they etched much slower.

5. Acknowledgements

This work was supported by the National Research, Development and Innovation Fund No. PD 121076, by the Hungarian Scientific Research Fund OTKA No. K 108366 and by the TAMOP 4.2.2.A-11/1/ KONV-2012-0036 project, which is co-financed by the European Union and European Social Fund.

The work was also supported by the GINOP 2.3.2-15-2016-00041 (co-financed by the European Union and the European Regional Development Fund).